\documentclass[fleqn,usenatbib]{mnras}

\usepackage[T1]{fontenc}
\usepackage{ae,aecompl}

\usepackage{graphicx}	% Including figure files
\usepackage{amsmath}	% Advanced maths commands
\usepackage{amssymb}	% Extra maths symbols

\usepackage{multirow}
\usepackage{booktabs}

\title[Geostrophic wind on oblate planets]{Geostrophic wind induced by latitudinal variation in gravitational acceleration on oblate planets}

\author[Haqq-Misra et al.]{Jacob Haqq-Misra$^{1,2}$\thanks{E-mail: jacob@bmsis.org}, Prabal Saxena$^{3}$, Eric T. Wolf$^{4}$, Ravi Kumar Kopparapu$^{1,2,3}$,\\
$^{1}$Blue Marble Space Institute of Science, 1001 4th Ave Suite 3201, Seattle, WA 98154, USA\\
$^{2}$NASA Astrobiology Institute's Virtual Planetary Laboratory, P.O. Box 351580, Seattle, WA 98195, USA\\
$^{3}$NASA Goddard Space Flight Center, 8800 Greenbelt Road, Mail Stop 699.0 Building 34, Greenbelt, MD 20771, USA\\
$^{4}$Department of Atmospheric and Oceanic Sciences, University of Colorado, Boulder, CO, 80309, USA}

\date{This version: 19 July 2016}

\pubyear{2016}

\begin{document}
\label{firstpage}
\pagerange{\pageref{firstpage}--\pageref{lastpage}}
\maketitle

\begin{abstract}
The population of known extrasolar planets includes giant and terrestrial planets that closely 
orbit their host star. Such planets experience significant tidal distortions that can force the planet 
into synchronous rotation. The combined effects of tidal deformation and centripetal acceleration 
induces significant asphericity in the shape of these planets, compared to the mild oblateness of Earth, 
with maximum gravitational acceleration at the poles. 
Here we show that this latitudinal variation in gravitational acceleration is relevant for modeling 
the climate of oblate planets including Jovian planets within the solar system, closely-orbiting 
hot Jupiters, and planets within the habitable zone of white dwarfs. We compare first- and third-order 
approximations for gravitational acceleration on an oblate spheroid and calculate the geostrophic 
wind that would result from this asphericity on a range of solar system planets and exoplanets.
Third-order variations in gravitational acceleration are negligible for Earth but become significant 
for Jupiter, Saturn, and Jovian exoplanets. 
This latitudinal variation in gravitational acceleration can be measured remotely, and the formalism presented 
here can be implemented for use in general circulation climate modeling studies of exoplanet atmospheres.
\end{abstract}

\begin{keywords}
planets and satellites: terrestrial planets -- planets and satellites: atmospheres -- hydrodynamics -- astrobiology
\end{keywords}

%%%%%%%%%%%%%%%%%%%%%%%%%%%%%%%%%%%%%%%%%%%%%%%%%%

%%%%%%%%%%%%%%%%% BODY OF PAPER %%%%%%%%%%%%%%%%%%

\section{INTRODUCTION}

The detection of thousands of extrasolar planets in the past two decades has included a significant number of planets orbiting closer to their host star than any planet in our solar system.   The discovery of such planets are disproportionate to their true occurrence rate largely due to the bias of the most prolific exoplanet detection methods: transit photometry and radial velocity.  The transit method generally achieves greater completeness for short period planets. This is due to the greater likelihood of observing a close-in planet that transits its host star over a discrete period of time.  An advantage of this bias is that this same effect also enables better characterization of the atmosphere and surface of these close-in exoplanets.  The higher transit frequency of the shortest period planets allows for signal-to-noise ratios that make these planets the most amenable to statistical characterization.  This has led to significant discoveries regarding the nature of short period gaseous \citep{2014Natur.505...69K, 2016Natur.529...59S} and solid planets \citep{2016Natur.532..207D}. 

These short period planets (with periods of $\lesssim$ several days) exist in an extreme stellar environment often characterized by high stellar radiation, planetary mass loss and extreme tidal forces, which are expected to occur in synchronous planetary orbits. One consequence of these extreme tidal forces and the fast rotation rates of these planets is that many of the closest of these worlds are likely to be aspherical in shape.  In fact, there have even been claimed detections of asphericity for WASP-12b \citep{2012ApJ...747...82C} and for a brown dwarf, Kepler 39b \citep{2014ApJ...796...67Z}.  However, asphericity is not unique to worlds outside our solar system.  Gas giants such as Jupiter and Saturn and a number of moons all possess a distorted shape more misshapen than the Earth.  An important consequence of this aspherical shape is that gravitational acceleration, $g$, for these bodies varies spatially on the surface of the planet. While such variations are small (but detectable) for the Earth, planets such as Jupiter and Saturn experience variations in $g$ more than an order of magnitude larger than the Earth.

This spatial variation in gravitational acceleration is typically assumed to be minimal enough to be ignored in climate modeling of the Earth and other worlds.  However, there are several notable exceptions to this.  The EPIC climate model \citep{2006Icar..182..259D} uses planetographic
latitude for oblate shape \citep{2014Icar..232..141P} in conjunction with latitude-variable gravity and also can be run using an orthogonal oblate-spherical coordinate system. This is in contrast to the bulk of climate models applied to atmospheres in the solar system (e.g. \citet{2010Icar..207..373L, 2012Icar..218..861F, 2014Icar..238..110G}), which do not include shape effects and associated effects on gravitational acceleration.  For exoplanets, nearly all climate models (to our knowledge) to date have also assumed spherical shape and constant gravitational acceleration on planetary surfaces.  Only the formalism developed by \citet{2015QJRMS.141.1646S, 2015QJRMS.141.2429S}, and its implementation in the Even  Newer  Dynamics  for General  Atmosphere  Modeling  of  the  Environment (ENDGame) dynamical core, has included these effects by developing appropriate non-spherical grids and equations of motion. Finally, while some atmospheric retrieval tools do incorporate shape effects \citep{2008JQSRT.109.1136I}, many tools used for exoplanet atmospheres still rely on assumptions of sphericity that are weakly held for close-in exoplanets.

Here we demonstrate the importance of planetary asphericity and associated variations in $g$ in modeling the climates of a number of solar system planets and exoplanets.  We first show that the larger observed and modeled asphericity of gas giants in our solar system and of hot Jupiters, respectively, necessitates a third-order approximation for varying $g$ versus traditional first-order approximations. We then discuss how this spatial variation in gravity results in asphericity-induced geostrophic wind components and variation in atmospheric layer thickness and lapse rate, which suggests that such shape effects are important for modeling of the atmospheres of gas giants and close in planets.  Finally, we discuss the observational implications of such effects and the prospects of remote detection.

\section{GRAVITATIONAL ACCELERATION ON AN OBLATE SPHEROID}

The shape of Earth and other planets are more accurately approximated as oblate spheroids, as a result of deformation due to rotation and asymmetries in mass distribution.
This implies that the constant value of gravitational acceleration, $g$, typically assumed for a spherical planet should instead vary 
as a function of latitude on an oblate spheroidal planet. A first-order approximation for $g$ was derived by French mathematician Alexis Claude Clairaut in 1743
using spherical harmonics. Although Clairaut's theorem is adequate for describing Earth's non-sphericity \citep{ramsey1940,jeffreys1976,Staniforth2015}, higher-order terms are required to more accurately 
describe the more extreme oblateness of the gas giants in our solar system and exoplanets such as hot Jupiters. In this section we summarize a third-order expansion of $g$ derived by \citet{Cook1959}
and compare these results with Clauraut's theorem for a range of planets.

To obtain an expression for gravitational acceleration on an oblate spheroid, we begin with the geometric definition
\begin{equation}
\frac{x^{2}+y^{2}}{a^{2}}+\frac{z^{2}}{c^{2}}=1,\label{eq:cartesian}
\end{equation}
where $a$ is equatorial radius, $c$ is polar radius, and
$(x,y,z)$ are Cartesian coordinates. We can express Eq. (\ref{eq:cartesian})
in terms of latitude ($\phi$), longitude ($\lambda$), and height
($z$) by applying the transformation $x=r\cos\lambda\cos\phi$, $y=r\sin\lambda\cos\phi$,
and $z=r\sin\phi$ to obtain the polar form of the equation: 
\begin{equation}
r^{2}\left(\frac{\cos^{2}\phi}{a^{2}}+\frac{\sin^{2}\phi}{a^{2}\left(1-\epsilon\right)^{2}}\right)=1,\label{eq:polar}
\end{equation}
where the non-dimensional parameter 
\begin{equation}
\epsilon=1 - \frac{c}{a}
\end{equation}
is the ellipticity of the spheroid. The mean radius of such a planet, $r_{m}$, is related to equatorial radius in terms of the ellipticity and semi-major
axis \citep{Cook1959} as: 
\begin{equation}
r_{m}=a\left(1-\frac{1}{3}\epsilon+\frac{1}{15}\epsilon^{2}+\frac{1}{35}\epsilon^{3}\right),\label{eq:meanradius}
\end{equation}
which allows us to calculate $a$ when $r_{m}$ is known from observations.
Along with the planetary rotation rate, $\Omega$, this allows us to define the non-dimensional parameter 
\begin{equation}
m=\frac{\Omega^{2}a^{3}}{GM} 
\end{equation}
as the ratio of centripetal to gravitational force at the equator. 
We also assume that the mean gravitational acceleration, $g_{m}$, can be 
known from observational constraints on planetary mass. 
We can then write a third-order expansion of $g$ on a spheroid as a function
of latitude \citep{Cook1959} as:
\begin{equation}
\begin{array}{c}
\frac{g}{g_{m}}=1+\left(\frac{5}{3}m-\frac{2}{3}\epsilon+\frac{10}{9}m^{2}+\frac{34}{63}\epsilon m-\frac{23}{63}\epsilon^{2}+\frac{20}{27}m^{3}+\frac{15}{14}\epsilon^{2}m+\right.\\
\left.\frac{46}{63}\epsilon m^{2}+\frac{421}{945}\epsilon^{3}\right)P_{2}(\cos\vartheta)+\\
\left(\frac{12}{35}\epsilon^{2}-\frac{4}{7}\epsilon m+\frac{982}{1155}\epsilon^{2}m-\frac{8}{27}\epsilon m^{2}+\frac{4}{11}\epsilon^{3}\right)P_{4}(\cos\vartheta)+\\
\frac{8}{231}\left(3\epsilon^{2}m-5\epsilon^{3}\right)P_{6}(\cos\vartheta),\label{eq:glat}
\end{array}
\end{equation}
where
\begin{equation}
\cos^{2}\vartheta=\left[1-\left(2\epsilon-\epsilon^{2}\right)\cos^{2}\phi-4\epsilon^{2}\sin^{2}\phi\cos^{2}\phi\right]\sin^{2}\phi
\end{equation}
and 
\[
\begin{array}{ccc}
P_{2}(x) & = & \frac{1}{2}\left(3x^{2}-1\right)\\
P_{4}(x) & = & \frac{1}{8}\left(35x^{4}-30x^{2}+3\right)\\
P_{6}(x) & = & \frac{1}{16}\left(231x^{6}-315x^{4}+105x^{2}-5\right)
\end{array}
\]
are Legendre polynomials. Eq. \ref{eq:glat} reduces to Clairaut's theorem when only the first-order terms in $\epsilon$ and $m$ are retained.

On Earth, expansion to $O(\epsilon)$ is usually sufficient to capture any significant departure from sphericity. However, other planetary
bodies show a larger deviation from sphericity than can be assumed for typical Earth-like conditions, requiring higher-order terms to be included. 
Our list of planets in Table \ref{table:planets} provides dimensional parameters $r_{m}$, $\Omega$, $M$, and $T_{c}$ (a characteristic atmospheric temperature) as well as non-dimensional parameters
$\epsilon$ and $m$ that are best-known estimates based upon observations. 
Using these parameters, we apply Eq. \ref{eq:glat} to examine the types of exoplanet atmospheres for which the latitudinal variation of gravitational acceleration becomes significant.
We first consider an idealized spherical planet with no asymmetry along any axis, followed by Earth, with a mild ellipticity of $\epsilon=0.003$. 
We also consider other terrestrial planets, such as Titan and 55 Cnc e, as well as Jupiter. We next examine 
the hot Jupiter planets WASP 12b, which resides in a close orbit around a G0-type star, and WASP 19b, which likewise occupies a close orbit
around a G8-type star. We last consider a hypothetical planet in orbit within the habitable zone of a white dwarf (WD) star; such planets
have not yet been discovered, but upcoming missions such as JWST should be able to detect these planets, if they exist.

An important note is that many of the close-in exoplanets listed here are actually likely tri-axial ellipsoidal in shape due to the superposition of tidal distortion onto the rotational distortion.  However, given that the polar axis and the orthogonal axis pointed toward the primary will still be the shortest and longest axes respectively, we use the oblate spheroidal approximation of shape for exoplanets in order to explore the effect and magnitude of the distorted shape on $g$.  The distortion of gaseous planets has been described in several ways, including the use of a Roche model to approximate the tri-axiality \citep{2011AJ....141...59B} and the use of a variational method and a simple polytropic assumption for the gaseous planet structure to produce an expression for asphericity \citep{2011A&A...528A..41L}. We use the latter method, which accounts for the dependence of structure on mass and radius as well as evolution with age, in order to estimate the asphericity of gaseous exoplanets. We model solid exoplanets in the same manner as \citet{2000ssd..book.....M} do for solar system bodies.  Putative solid exoplanets are assigned an effective rigidity ($\mu$), which relates the elastic strength of the planet to its gravity.  We assume a value of $\mu$ of $\sim100$ GPa, about the mean value for the Earth \citep{2005soea.book.....F}.  Different values such as a higher rigidity equivalent to the Earth's core still yield significant asphericities.  Additionally, given that current hypotheses regarding these planets range from rigid Iron-core hot Jupiter remnants to planets which may significantly be less rigid due to partial or whole body melting, any assumed rigidity value would be somewhat arbitrary.

\begin{table*}
\centering
\caption{A list of planets and associated parameters analyzed in this study.}
\label{table:planets}
\begin{tabular}{c||cccc|cc}
Planet & $r_{m}$ & $\Omega$ & $M$ & $T_{c}$ & $\epsilon$ & $m$\tabularnewline
\hline 
\hline 
Sphere & $6.371\times10^{6}\text{ m}$ & $7.27\times10^{-5}\text{ s}^{-1}$ & $5.972\times10^{24}\text{ kg}$ & $250\text{ K}$ & $0$ & $0$\tabularnewline
Earth & $6.371\times10^{6}\text{ m}$ & $7.27\times10^{-5}\text{ s}^{-1}$ & $5.972\times10^{24}\text{ kg}$ & $250\text{ K}$ & $0.003$ & $0.003$\tabularnewline
Titan & $2.575\times10^{6}\text{ m}$ & $4.57\times10^{-6}\text{ s}^{-1}$ & $1.345\times10^{23}\text{ kg}$ & $90\text{ K}$ & $0.0002$ & $0.00004$\tabularnewline
Jupiter & $6.991\times10^{7}\text{ m}$ & $1.75\times10^{-4}\text{ s}^{-1}$ & $1.898\times10^{27}\text{ kg}$ & $150\text{ K}$ & $0.06$ & $0.089$\tabularnewline
55 Cnc e & $1.274\times10^{7}\text{ m}$ & $9.87\times10^{-5}\text{ s}^{-1}$ & $5.135\times10^{25}\text{ kg}$ & $2500\text{ K}$ & $0.016$ & $0.0063$\tabularnewline
WASP 19b & $9.690\times10^{7}\text{ m}$ & $9.22\times10^{-5}\text{ s}^{-1}$ & $2.217\times10^{27}\text{ kg}$ & $2500\text{ K}$ & $0.13$ & $0.075$\tabularnewline
WASP 12b & $1.258\times10^{8}\text{ m}$ & $6.71\times10^{-5}\text{ s}^{-1}$ & $2.657\times10^{27}\text{ kg}$ & $2500\text{ K}$ & $0.12$ & $0.068$\tabularnewline
WD Planet & $6.371\times10^{6}\text{ m}$ & $1.75\times10^{-4}\text{ s}^{-1}$ & $5.972\times10^{24}\text{ kg}$ & $250\text{ K}$ & $0.02$ & $0.02$\tabularnewline
\end{tabular}
\end{table*}

\begin{figure*}
\centering
\includegraphics[width=1.\linewidth]{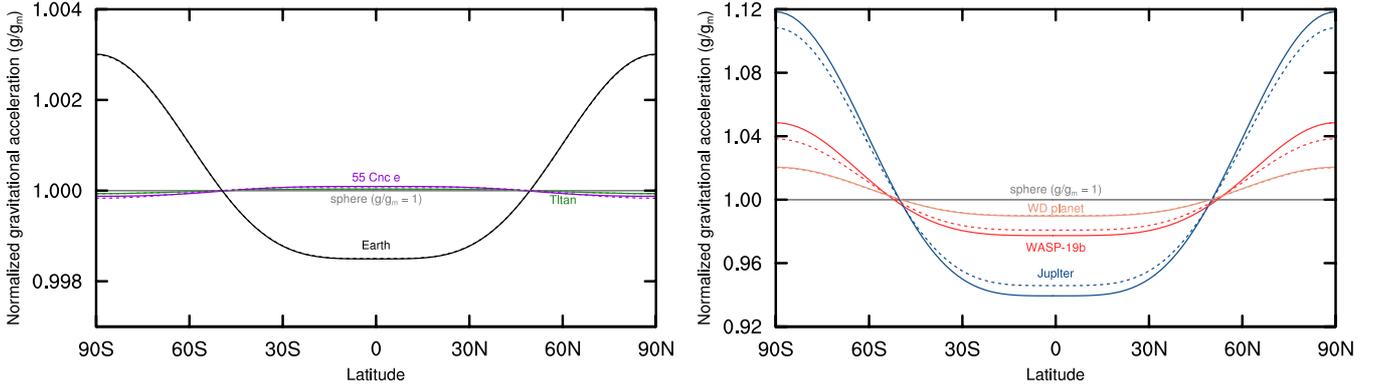}
\caption{Normalized latitudnal variation of gravitational acceleration, $g$, for 
select terrestrial planets (top panel) and gas giants (bottom panel).
Solid lines show third-order calculations of $g$ \citep{Cook1959}, while dashed lines 
show first-order calculations using Clairaut's theorem.}
\label{fig:gravity}
\end{figure*}

Figure \ref{fig:gravity} shows the normalized latitudinal variation of $g$ for select terrestrial planets and gas giants. 
Solid curves show a third-order approximation for $g$ \citep{Cook1959}, while dashed curves show the first-order approximation using Clairaut's theorem.
The effect of non-sphericity on Earth contributes less than a percent toward latitudinal variation in $g$, with a maximum value at the poles and minimum 
at the equator (left panel). Although Earth bulges slightly with its maximal mass distribution at the equator, 
this results in the poles being closer to Earth's center, which results in maximum $g$ at the poles.
By contrast, Titan and 55 Cnc e both show a slightly greater $g$ at the equator and a minimum at the poles. Gas giants 
Jupiter and WASP 12b, as well as our hypothetical WD planet, show a maximum in $g$ at the poles, with several percent in latitudinal variation due to oblateness (right panel). 
Earth and other terrestrial plants show little difference between the first- and third-order approximations of $g$, but gas giants show a greater deviation 
between these two forms. 
In addition, the top panel of Figure \ref{fig:planets} shows a plot of the pole to equator variation in $g$ (with the highest values at the pole) for the most distorted solid and gaseous exoplanets, the solar system gas giants and the Earth.  Synchronous rotation is assumed for the modeled exoplanets.  The most distorted gaseous exoplanets experience gravity variations approaching 5\% while their solar system counterparts have variation approaching 15 and 25\%.  The closest in solid exoplanets can experience similar variations.
These calculations suggest that Earth-based climate models probably have little to gain from including latitudinal 
variation in $g$, but this effect becomes more significant in gas giant and exoplanet atmospheres.

%**********************************************
%Fig. 1
%**********************************************

\section{GEOSTROPHIC WIND DUE TO OBLATENESS}

We next examine the effect of latitudinally-varying gravitational acceleration on geostrophic wind.
Geostrophic balance is a reduction of the full Navier-Stokes equations of motion into a diagnostic 
relationship between the pressure gradient and Coriolis forces. The resulting theoretical 
wind pattern that would arise from balance between only these two forces is known as 
the geostrophic wind. Geostrophic analysis provides an approximation of atmospheric motion
at a given state, but as a diagnostic relationship it cannot make any predictions about the 
future evolution of weather systems. For our purposes, the geostrophic wind will provide 
insight into the magnitude of contribution that oblateness can exert on climate.

Although we already have an expression for $g$ as a function of latitude (Eq. \ref{eq:glat}), we cannot
apply this expression to the equations of motion without first transforming our underlying coordinate system.
\citet{white2005} demonstrated that including latitudinally-varying $g$ with a spherical geometical approximation
results in spurious and unphysical sources of vorticity. Ellipsoidal coordinate systems avoid this problem, 
but \citet{white2008} argue that the typical use of confocal oblate spheroidal coordinates suffers from a different 
problem; namely, the separation of ellipsoids with height incorrectly leads to a greater value of $g$ at the equator 
than the poles for Earth (contrary to observations). A better solution, that avoids both of these problems, is to use a coordinate system 
of similar oblate spheroids \citep{white2008} for the geometry of the planet and the equations of motion.

\citet{Staniforth2015} derive the shallow water equations using a similar oblate spheroid coordinate system, which 
can be applied to the planets considered here. We reduce their system of equations in geographic ellipsoidal coordinates (eqs.
58 and 59 in \citet{Staniforth2015}) into a corresponding set of equations for the geostrophic wind:
\begin{equation}
v_{g}=-\frac{1}{2\Omega\sin\phi}\frac{r_{ac}}{a^{2}\cos\phi}\frac{\partial\Phi}{\partial\lambda},\label{eq:vwind}
\end{equation}
\begin{equation}
u_{g}=\frac{1}{2\Omega\sin\phi}\frac{r_{ac}^{3}}{a^{2}c^{2}}\frac{\partial\Phi}{\partial\phi},\label{eq:uwind}
\end{equation}
where 
\begin{equation}
r_{ac}=\left(a^{2}\cos^{2}\phi+c^{2}\sin^{2}\phi\right)^{\frac{1}{2}}.
\end{equation}
Here the geopotential, $\Phi$, is written as the product of gravitational
acceleration and height: $\Phi=g(\phi)H$. 

Taken alone, the equations
of geostrophic balance (\ref{eq:vwind}) and (\ref{eq:uwind}) are
degenerate and cannot yield a steady-state solution without additional
information \citep{gill1982}. We assume for simplicity that $H=H_{0}=RT_{c}/g_{m}$,
with $T_{c}=250$ K for Earth, $T_{c}=2525$ K for hot Jupiters, and other 
values of $T_{c}$ given in Table \ref{table:planets}, which represents atmospheric height 
through a scaling relationship and assumes a surface with no topography. This leads to a steady-state
solution of the geostrophic wind on geographic ellipsoidal coordinates
with latidudinal-varying gravity:
\begin{equation}
v_{g}=0 ,\label{eq:vwind-1}
\end{equation}
\begin{equation}
\begin{array}{c}
u_{g}=\frac{RT_{c}}{2\Omega\sin\phi}\frac{r_{ac}^{3}}{a^{2}c^{2}}\left\{ \left(\frac{5}{3}m-\frac{2}{3}\epsilon+\frac{10}{9}m^{2}+\frac{34}{63}\epsilon m-\frac{23}{63}\epsilon^{2}+\frac{20}{27}m^{3}+\right.\right.\\
\left.\frac{15}{14}\epsilon^{2}m+\frac{46}{63}\epsilon m^{2}+\frac{421}{945}\epsilon^{3}\right)P_{2}^{'}(\cos\vartheta)+\\
\left(\frac{12}{35}\epsilon^{2}-\frac{4}{7}\epsilon m+\frac{982}{1155}\epsilon^{2}m-\frac{8}{27}\epsilon m^{2}+\frac{4}{11}\epsilon^{3}\right)P_{4}^{'}(\cos\vartheta)+\\
\\
\left.\frac{8}{231}\left(3\epsilon^{2}m-5\epsilon^{3}\right)P_{6}^{'}(\cos\vartheta)\right\} ,
\end{array}\label{eq:uwind-1}
\end{equation}
where
\begin{equation}
\begin{array}{c}
P_{2}^{'}(\cos\vartheta)=3\sin\phi\cos\phi\left[\epsilon^{2}\left(4\sin^{4}\phi-8\sin^{2}\phi\cos^{2}\phi+\right.\right.\\
\left.\left.2\cos^{2}\phi-1\right)+2\epsilon\left(1-2\cos^{2}\phi\right)+1\right]
\end{array},
\end{equation}
\begin{equation}
\begin{array}{c}
P_{4}^{'}(\cos\vartheta)=\frac{5}{2}\sin\phi\cos\phi\left[28\epsilon^{3}\sin^{2}\phi\cos^{2}\phi\left(1-\right.\right.\\
\left.4\sin^{4}\phi-2\cos^{2}\phi\right)+\\
3\epsilon^{2}\left(3\sin^{4}\phi-28\sin^{4}\phi\cos^{2}\phi+56\sin^{4}\phi\cos^{4}\phi+\right.\\
\left.28\sin^{2}\cos^{4}\phi-6\sin^{2}\phi\cos^{2}\phi-2\cos^{2}\phi+1\right)+\\
2\epsilon\left(7\sin^{4}\phi-14\sin^{2}\cos^{2}\phi+6\cos^{2}\phi-3\right)-\\
\left.7\cos^{2}\phi+4\right],
\end{array}
\end{equation}
\begin{equation}
\begin{array}{c}
P_{6}^{'}(\cos\vartheta)=\frac{-21}{8}\sin\phi\cos\phi\left[4\epsilon^{3}\sin^{2}\phi\cos^{2}\phi\left(88\sin^{6}\phi+\right.\right.\\
34\sin^{4}\phi-506\sin^{4}\phi\cos^{2}\phi+15\sin^{2}\phi\cos^{2}\phi+\\
\left.330\sin^{2}\phi\cos^{2}\phi-44\cos^{4}\phi+28\cos^{2}\phi-14\right)+\\
\epsilon^{2}\left(-28\sin^{6}\phi+3\sin^{4}\phi+440\sin^{6}\phi\cos^{2}\phi-\right.\\
46\sin^{4}\phi\cos^{2}\phi-792\sin^{4}\phi\cos^{4}\phi+2\sin^{2}\phi\cos^{2}\phi+\\
130\sin^{2}\phi\cos^{4}\phi+88\sin^{2}\phi\cos^{6}\phi-\\
\left.4\cos^{4}\phi-2\cos^{2}\phi+1\right)+\\
2\epsilon\left(-7\sin^{4}\phi+66\sin^{4}\phi\cos^{2}\phi+6\sin^{2}\phi\cos^{2}\phi-\right.\\
\left.66\sin^{2}\phi\cos^{4}\phi+4\cos^{4}\phi+2\cos^{2}\phi-1\right)-\\
\left.11\sin^{4}\phi+22\sin^{2}\phi\cos^{2}\phi-8\cos^{2}\phi+3\right],
\end{array}
\end{equation}
are the derivates of the Legendre polynomials with repect to latitude,
with terms $O(\epsilon^{3})$ and lower retained.
We examine this third order expression for zonal geostrophic wind described by Eq. (\ref{eq:uwind-1}) as well as 
a first-order approximation when $g$ follows Clairaut's theorem.

%**********************************************
%Fig. 2
\begin{figure*}
\centering
\includegraphics[width=1.\linewidth]{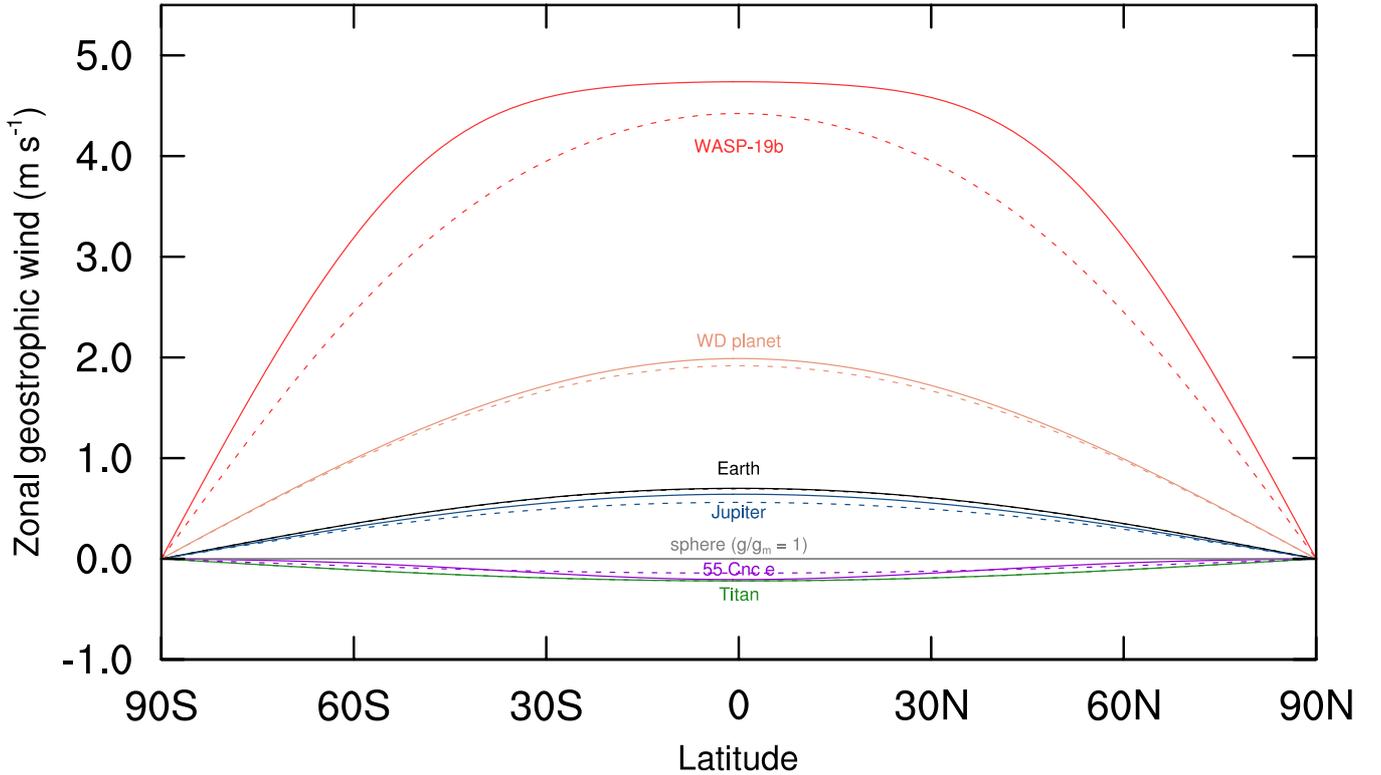}
\caption{Zonal geostrophic wind, $u_{g}$, induced by latitudinal variations 
in gravitational acceleration due to asphericity. 
Solid lines show third-order calculations of $g$ \citep{Cook1959}, while dashed lines
show first-order calculations using Clairaut's theorem.}
\label{fig:wind}
\end{figure*}
%**********************************************

Figure \ref{fig:wind} shows the zonal geostrophic wind as a function of latitude for select terrestrial planets and gas giants.
Earth, Jupiter, WASP 19b and the theoretical WD planet all show an increase in $u_g$ due to latitudinally-varying $g$, with a maximum
at the equator and minimum at the poles. This effect is only about $0.5$\,m\,s$^{-1}$ for Earth and Jupiter, but the geostrophic zonal 
wind increases to a maximum of about 2\,m\,s$^{-1}$ for the WD planet and 5\,m\,s$^{-1}$ for WASP 19b. Conversely, Titan 
and 55 Cnc show a decrease in wind speeds of up to about $-0.5$\,m\,s$^{-1}$. Third-order effects appear most pronounced for 
WASP 19b, as this closely orbiting hot Jupiter experiences an extreme degree of distortion due to tidal forces and its rapid rotation. Jupiter and the 
WD planet show some deviation between third-order and first-order approximations, while terrestrial planets show little to no variation
between the two approaches to $g$. 

\begin{figure}
  \centering
  \includegraphics[scale=0.5]{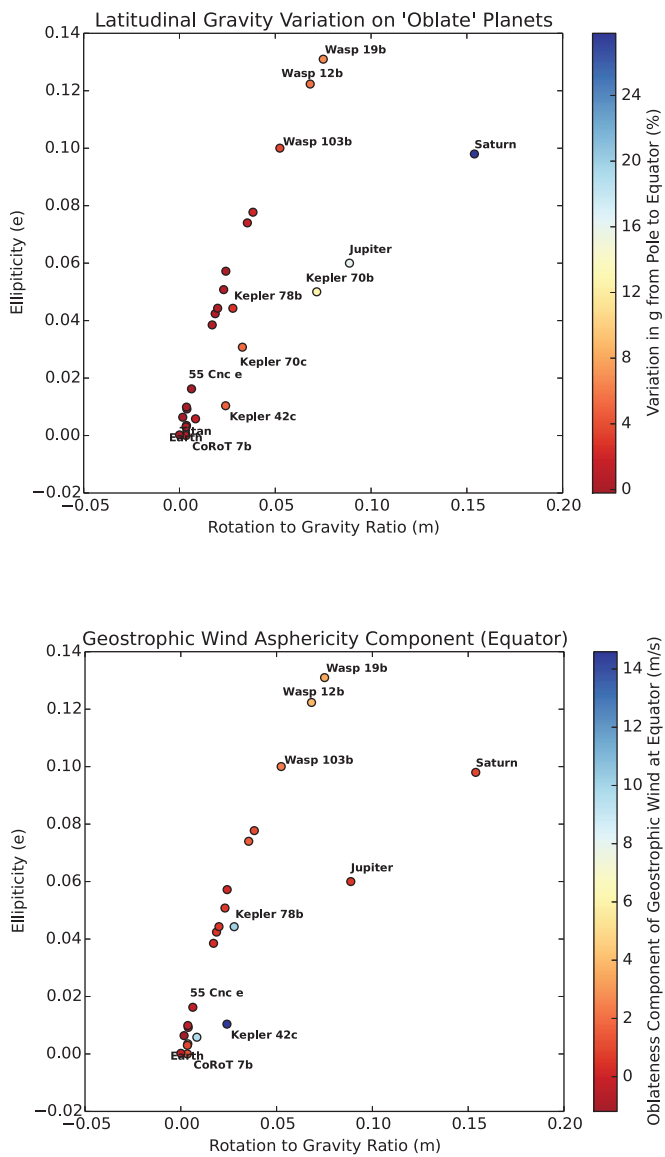}
  \caption{Latitudinal variation of gravitational acceleration, $g$, for 
the most distorted solar system planets and the most distorted solid and gaseous exoplanets (top figure).   Asphericity components of the zonal geostrophic wind at the equator for the same planets (bottom figure).  Asphericities for exoplanets are modeled using \citep{2011A&A...528A..41L, 2015MNRAS.446.4271S}. }\hspace{-0.8 cm}
  \label{fig:planets}
\end{figure}

The bottom panel of Figure \ref{fig:planets} shows the asphericity component of the zonal geostrophic wind at the equator for the selected planets.  In this case, planetary equilibrium temperatures were used.  The rapidly rotating scorched putative rocky exoplanets exhibit the largest asphericity components to the zonal geostrophic wind.  In fact, Kepler 70b and c, planets on sub 9 hour orbits around a subdwarf B star \citep{2011Natur.480..496C}, are excluded from the plot because their wind components exceed 100\,m\,s$^{-1}$.
These results suggest that the atmospheres of hot Jupiters and planets orbiting in the habitable 
zone of white dwarf stars will include contributions due to asphericity. Any attempt to represent this phenomenon for hot Jupiters
in climate models should include higher-order terms to accurately represent the contribution of latitudinally-varying graviational
acceleration to the equatorial geostrophic wind. 

\section{ATMOSPHERIC THICKNESS ON AN OBLATE SPHEROID}
We also consider the effect on the observed thickness of the atmosphere due to latitudinally-varying gravitational acceleration. 
Hydrostatic balance describes the equilibrium in the atmosphere between the downward force of 
gravity and the pressure gradient force. Although the actual atmosphere is not exactly in hydrostatic balance, 
this still provides a good approximation for many scales of motion. In this section we examine hydrostatic balance, and the related 
hypsometric equation, on oblate planets.

\begin{figure}
\centering
\includegraphics[scale=0.42]{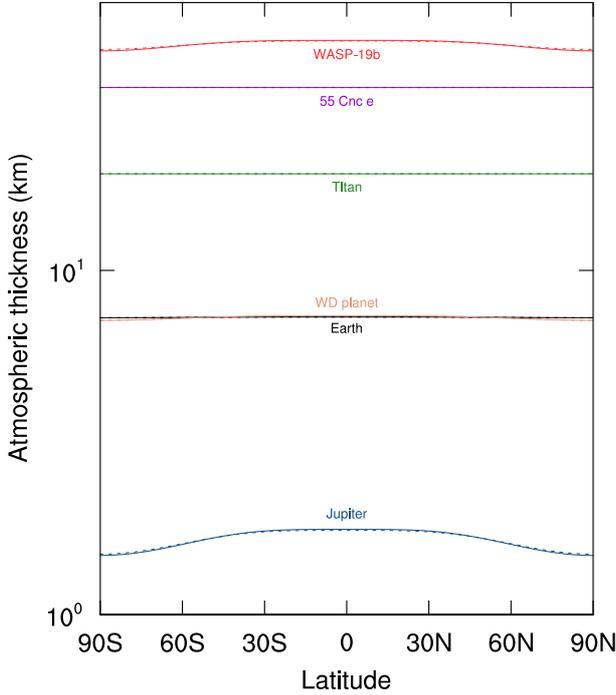}
\caption{Atmospheric thickness, $Z$, at one scale height resulting from latitudinal variations 
in gravitational acceleration due to oblateness. 
Solid lines show third-order calculations of $g$ \citep{Cook1959}, while dashed lines
show first-order calculations using Clairaut's theorem.}
\label{fig:thickness}
\end{figure}

The condition of hydrostatic balance can be written as:
\begin{equation}
\frac{dp}{dz}=-\rho g,\label{eq:hydrostatic}
\end{equation}
where $\rho$ is density and $p$ is pressure. Integrating Eq. (\ref{eq:hydrostatic}) from a height $z$ to the
top of the atmosphere gives 
\begin{equation}
p(z,\phi)=\int_{z}^{\infty}\rho gdz.\label{eq:hydrostaticintegral}
\end{equation}
This relationship shows that atmospheric pressure on an oblate spheroidal planet depends on both height and latitude. 
Although this equation provides a diagnostic tool for assessing atmospheric mass, 
climate models and analysis typically represent height using geopotential, $\Phi\equiv\nabla g$,
rather than pressure. We can rewrite the hydrostatic equation (\ref{eq:hydrostatic})
in terms of geopotential as: 
\begin{equation}
gdz=d\Phi=-RTd\ln p,\label{eq:gdz}
\end{equation}
where we have invoked the ideal gas law $p=\rho RT$. We can then
vertically integrate Eq. (\ref{eq:gdz}) to obtain an expression for
the thickness of the atmosphere, which is known as the hypsometric equation:
\begin{equation}
\Phi(z_{2},\phi)-\Phi(z_{1},\phi)=g(H_{2}-H_{1})=R\int_{p_{2}}^{p_{1}}Td\ln p.\label{eq:hypsometric}
\end{equation}
We define the difference in height $H_{2}-H_{1}$ as the thickness, $Z$, so that 
\begin{equation}
Z=H_{2}-H_{1}=\frac{R}{g}\int_{p_{2}}^{p_{1}}Td\ln p.\label{eq:thickness}
\end{equation}
If we assume a mean scale height $H_{0}=RT_{c}/g_{m}$ as before,
and we integrate from the surface to the top of atmosphere, then Eq.
(\ref{eq:thickness}) becomes
\begin{equation}
Z=-H_{0}\left(\frac{g_{m}}{g}\right)\ln\frac{p}{p_{0}},\label{eq:thickness2}
\end{equation}
which shows that $Z$ depends upon latitude on
oblate planets. If we solve Eq. (\ref{eq:thickness2}) for $p$, then we obtain an expression 
for the exoponential decrease of pressure with height in an isothermal 
atmosphere:
\begin{equation}
p=p_{0}\exp\left[\frac{-Z}{H_0}\left(\frac{g}{g_{m}}\right)\right].\label{eq:pthickness}
\end{equation}
The term $({Z}/{H})({g}/{g_{m}})$ is scale height of the atmosphere at which pressure deceases by a factor of $e^{-1}$.
Figure \ref{fig:thickness} shows the latitudinal dependence of atmospheric thickness $Z$ for an assumed
pressure over a single scale height. Earth, Titan, and 55 Cnc all show flat latitudinal profiles with no contribution to thickness from oblatness. By contrast, Jupiter and WASP-19b both show greater thickness at the
equator and reduced thickness at the poles, with near agreement between the use of first-order (dashed curve) and 
third-order (solid curve) approximations of $g$. The hypothetical WD planet also shows similar effects, although 
the contribution to thickness from non-sphericity is much more significant for giant planets.

%**********************************************
%Fig. 3
%**********************************************

Earth-based observations typically assume that $Z$ is independent of latitude, which allows for 
curvature effects to be mathematically simplified. For example, the quasi-geostrophic approximation invokes 
the beta-plane approximation, which assumes the Coriolis parameter is set to vary linearly in reference to a fixed value. Although the beta-plane approximation is operationally successful on a mildly oblate planet like 
Earth, such simplifications would be inadequate for describing large-scale motion on planets with a more 
extreme departure from asphericity, particularly on giant planets. Analytic approximations to diagnosing large-scale atmospheric phenomenon on Earth may therefore be limited in their application to aspherical exoplanets.

\section{Thermodynamics and Lapse Rates}

We finally consider the contribution of oblateness to the thermodynamics of a planetary atmosphere. 
For a dry parcel of air undergoing an adiabatic processes (i.e., a reversible processes in which no energy is
exchanged with the environment), the first law of thermodynamics can be expressed as 
\begin{equation}
\theta=T\left(\frac{p_{0}}{p}\right)^{R/c_{p}},\label{eq:ptemp}
\end{equation}
where $\theta$ is known as potential temperature. Potential temperature is the temperature that a parcel 
of air would have if it were expanded or compressed adiabatically to surface pressure $p_0$. Aside 
from regions of precipitation, large-scale atmospheric motions can generally be approximated 
as adiabatic. 

Lapse rate refers to the rate of decrease in temperature with height. We can express lapse rate 
by first taking the derivative of Eq. (\ref{eq:ptemp}) with respect to height and then substituting from 
the hydrostatic equation (\ref{eq:hydrostatic}) and ideal gas law to obtain 
\begin{equation}
\frac{T}{\theta}\frac{\partial\theta}{\partial z}=\frac{\partial T}{\partial z}+\frac{g}{c_{p}}.\label{eq:lapse2}
\end{equation}
If we assume that the atmosphere can be approximated as adiabatic, then potential temperature is constant with height. This reduces Eq. (\ref{eq:lapse2}) to
\begin{equation}
-\frac{\partial T}{\partial z}=\frac{g}{c_{p}}\equiv\Gamma_{d}(\phi),\label{eq:lapse}
\end{equation}
where $\Gamma_{d}$ is known as the dry adiabatic lapse rate. Typically for Earth-centric studies, 
$\Gamma_{d}$ is assumed to be a constant 9.8\,K\,km$^{-1}$, but for oblate planets 
Eq. (\ref{eq:lapse}) shows that $\Gamma_{d}$ also depends on latitude. 

Inspection of Figure \ref{fig:gravity} reveals the planets most affected by the latitudinal dependence 
of lapse rate. On Earth this effect is about one part in one thousand from equator to midlatitudes, and the 
effect is even less for Tital and 55 Cnce. By contrast, Jupiter shows nearly a 20\% change in $g$, and thus 
$\Gamma_{d}$, from equator to pole, and WASP-12b likewise shows a similar change of about 10\%. 
Such planets would show a lower lapse rate in the tropics and a higher lapse rate at the poles. 
(Although we are focusing on the dry lapse rate $\Gamma_{d}$, the moist adaibatic lapse rate 
is also proporational to $g$, so this discussion also applies to moist atmospheres.) 

This latitudinal dependence of lapse rate would feedback upon atmospheric thickness (as described by Eq. (\ref{eq:thickness})) by modifying the temperature structure.  In turn, these temperature dependent changes to atmospheric thickness (and thus the geopotential) would then feedback upon the zonal component of the geostrophic wind (Eq. \ref{eq:uwind}).  These combined effects further suggest that the large-scale atmospheric motion on oblate planets may be inadequately described by dynamical theory solved on a sphere.  Additionally, one may expect cloud-forming regions will be affected, because the environmental lapse rate largely determines the lifting condensation level in the atmosphere.  Predicting the effect of oblateness on cloud formation would require the use of general circulation models.

%**********************************************
%Fig. 4
%**********************************************

\section{IMPLICATIONS FOR OBSERVATIONS}

\begin{figure}
\centering
\includegraphics[scale=0.5]{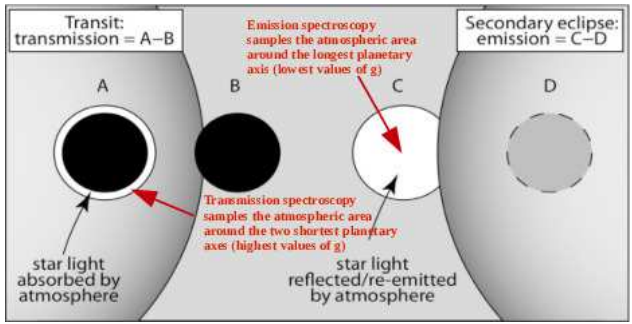}
\caption{{The tri-axial shape of a distorted planet would result in transmission and emission spectroscopy sampling parts of an atmosphere with the most extreme opposing values of g.  For significantly distorted planets, this variation in g could be akin to sampling entirely different atmospheres.  Adapted from Image by Phillip Amrein. }}
\label{fig:transem}
\end{figure}

Asphericity-induced spatial variations of $g$ have the potential to influence atmospheric observables in a number of ways.  The geometric configuration taken advantage of by transit spectroscopy is one of the most clear examples of this.  For edge-on transiting planets in near circular orbits, transmission spectroscopy probes the atmospheric annulus of the planet that extends from pole to pole through the terminator.  Meanwhile, emission spectroscopy allows for the analysis of the dayside atmosphere of a planet. Given the likely tri-axial ellipsoid shape of most of the close-in planets in this study, these two separate regions of the planets that are being observed correspond to two parts of the planet where the difference in values of $g$ is a maximum.  This can result in disparate or even conflicting interpretations of the atmosphere (Figure \ref{fig:transem}).

More generally for both oblate spheroidal and tri-axial ellipsoid bodies, since mixing ratios and vertical mixing scale linearly with $g$, such variations can effect chemistry spatially. The effect of these variations in $g$ can also influence pressure layer thickness of the atmosphere and atmospheric density. For synchronously rotating exoplanets, horizontally propagating gravity waves may dominate day-to-night heat transport \citep{2013ApJ...776..134P}, fundamentally altering not only the observed temperature gradients, but also fundamentally controlling the atmospheric conditions found at the terminator, the region which is observed using transmission spectroscopy. Evidence of the impact of variations in wind velocity at the terminator can be detected in transiting planets through the measurement of Doppler broadening and shifting of individual spectral features using high-resolution spectroscopy; such velocity-broadened spectral features have already been identified in the directly imaged planets \citep{snellen2014} and may soon be detectable in transiting planets \citep{rauscher2014}.

In addition to the effect on spectroscopy and retrieval, such asphericity-induced variation can also confound interpretation of dynamical features in the atmosphere that may be visible in observations of Jupiter and Saturn and also in photometric mapping studies of exoplanets.  Asphericity-induced components of zonal equatorial geostrophic winds may approach $>$10\,m\,s$^{-1}$ and $\sim$\,5\,m\,s$^{-1}$ for solid and gaseous planets respectively. While this is small compared to the $>$1\,km\,s$^{-1}$ winds in the upper atmospheres of hot Jupiters \citep{2009ApJ...700..887M}, it is a non-trivial factor compared to some of the winds modeled in the deep atmosphere, and outside the jet region. \citep{2016ApJ...821....9K}. Additionally, while these models assume an oblate spheroidal shape that may apply well to solar system gas giants and brown dwarfs, close-in exoplanets are likely to be triaxial ellipsoids.  In this case, there would be significant polar and equatorial oriented geostrophic wind components.  These components are likely to be more important relatively to other equatorial and polar flows and may effect global flow patterns and positions of features such as hotspots and clouds.

%**********************************************
%Fig. 5
%**********************************************

\section{CONCLUSIONS}

Due to the close proximity to their host star and their short, synchronized orbits, some of the planets most statistically amenable to characterization are also expected to be the most tidally and rotationally deformed in shape.  Planets in our solar system such as Jupiter and Saturn are also distorted to a significant degree and are increasingly becoming a focus of observations and missions (such as JUNO \citep{2007AcAau..61..932M}).  Given the heightened interest in these aspherical worlds, corresponding climate and atmosphere studies should at least consider the effects of aspherical shape on local values of $g$ and the subsequent influence on atmospheric structure, dynamics and chemistry.  

In general, such planets are likely to experience spatial variation in $g$ significantly larger than the Earth and may experience shape driven atmospheric dynamics effects.  Such effects may influence the state of the atmosphere and consequently may be important to observations of these worlds.   While the oblate spheroidal shape assumed for this study may be appropriate for solar system gas giants and brown dwarfs, there is further need to understand how $g$ would vary on tri-axial ellipsoid-shaped bodies that are more representative of close-in exoplanets.  Regardless of the exact shape however, the larger departure from sphericity of these worlds should drive the need to better understand the atmospheric effects of asphericity.

\section*{ACKNOWLEDGEMENTS}
The authors thank Avi Mandell and \'{E}ric H\'{e}brard for discussions and comments that improved this study.
J.H., E.T.W., and R.K.K. acknowledge funding from the NASA Habitable Worlds program under award NNX16AB61G.  P.S. was supported by
an appointment to the NASA Postdoctoral Program at Goddard Space Flight Center, administered by Universities Space Research Association through a contract with NASA.
Any opinions, findings, and conclusions or recommendations expressed in this material are those of the authors and do not necessarily reflect the views of NASA.

%%%%%%%%%%%%%%%%%%%%%%%%%%%%%%%%%%%%%%%%%%%%%%%%%%



\bibliographystyle{mnras}
\bibliography{oblatesphere_ms} % if your bibtex file is called example.bib

%%%%%%%%%%%%%%%%%%%%%%%%%%%%%%%%%%%%%%%%%%%%%%%%%%

% Don't change these lines
\bsp	% typesetting comment
\label{lastpage}
\end{document}